\documentclass{emulateapj}
\newcommand{\boldsymbol}[1]{\mbox{\boldmath{$#1$}}}

\shorttitle{Evolution of Magnetic Turbulence in Shocks} 
\shortauthors{Chang, Spitkovsky, \& Arons}

\begin{document}

\title{Long Term Evolution of Magnetic Turbulence in Relativistic Collisionless Shocks: Electron-Positron Plasmas} 

\author{Philip Chang\altaffilmark{1,2,3}, Anatoly Spitkovsky\altaffilmark{4}, and Jonathan Arons\altaffilmark{1,2,5,6}}

\altaffiltext{1} {Department of Astronomy, Campbell Hall, University
  of California, Berkeley, CA 94720; pchang@astro.berkeley.edu,
 arons@astro.berkeley.edu}
\altaffiltext{2} {Theoretical Astrophysics Center} 
\altaffiltext{3} {Miller Institute for Basic Research}
\altaffiltext{4} {Department of Astrophysical Sciences, Peyton Hall,
Princeton University, Princeton, NJ: anatoly@astro.princeton.edu}
\altaffiltext{5} {Department of Physics, LeConte Hall, University
  of California, Berkeley, CA 94720}
\altaffiltext{6}{Kavli Institute for Particle Astrophysics and Cosmology, Stanford
University}

\begin{abstract}
  We study the long term evolution of magnetic fields generated by a
  collisionless relativistic $e^+e^-$ shock which is initially
  unmagnetized.  Our 2D particle-in-cell numerical simulations show
  that downstream of such a Weibel-mediated shock, particle
  distributions are close to isotropic, relativistic Maxwellians, and
  the magnetic turbulence is highly intermittent spatially. The
  non-propagating magnetic fields in the turbulence form relatively isolated regions
  with transverse dimension $\sim 10-20$ skin depths. These structures
  decay in amplitude, with little sign of downstream merging.  The
  fields start with magnetic energy density $\sim (0.1-0.2)$ of the
  upstream kinetic energy within the shock transition, but rapid
  downstream decay drives the fields to much smaller values, below
  $10^{-3}$ of equipartition after $\sim 10^3$ skin depths.

%  At late times in the simulations, which last for
%  $\sim 4000 \omega_p^{-1}$ and follow the motions of $15 \time 10^9$
%  particles, where $\omega_p$ is the relativistic plasma frequency
%  based on upstream parameters, the total magnetic energy, averaged
%  over the transverse dimension, declines in proportion to $t^{-1}$.
%  Noise due to the finite number of particles limits following this
%  decay to magnetic energy densities greater than $0.001 nkT$.
  
  In an attempt to construct a theory that follows field decay to
  these smaller values, we explore the hypothesis that the observed
  damping is a variant of Landau damping in an unmagnetized plasma.
  The model is based on the small value of the downstream magnetic
  energy density, which
  suggests that particle orbits are only weakly perturbed from
  straight line motion, if the turbulence is homogeneous.  Using
  linear kinetic theory applied to electromagnetic fields in an
  isotropic, relativistic Maxwellian plasma, we find a simple analytic
  form for the damping rates, $\gamma_k$, in two and three dimensions
  for small amplitude, subluminous electromagnetic fields.  We find
  that magnetic energy does damp due to phase mixing of current
  carrying particles as $(\omega_p t)^{-q}$ with $q \sim 1$. This
  overall decay compares well to that found in simulations, since it
  depends primarily on the longest wavelength modes, $kc/\omega_p \ll
  1$.  However, the theory predicts overly rapid damping of short
  wavelength modes.  We speculate that magnetic trapping of a
  substantial fraction of the particles within the highly spatially
  intermittent downstream magnetic structures may be the origin of
  this discrepancy.  In addition, trapping may form the basis for
  MHD-like behavior, permitting a small fraction of the initial
  magnetic energy to persist for times much greater than have been
  followed in the simulations.

  We briefly speculate on other physical processes, which depend on
  the presence of suprathermal particles, that may cause the
  generation of longer wavelength magnetic fields that create a
  magnetized plasma ($kr_{Larmor} \ll 1$), in which the damping is not
  as fast.  However, absent such additional physical processes, we
  conclude that initially unmagnetized relativistic shocks in
  electron-positron plasmas are unable to form persistent downstream
  magnetic fields.  These results put interesting constraints on
  synchrotron models for the prompt and afterglow emission from GRBs.
  We also comment on the relevance of these results for relativistic
  shocks in electron-ion plasmas.

% Because
%  the downstream magnetic turbulence actually is highly intermittent
%  in space, magnetic trapping of a substantial fraction of the
%  particles within the magnetic structures occurs and may be the
%  origin of the discrepancy between the theory and the late time
%  damping observed numerically.  We speculate that trapping may allow
%  persistence of a small fraction of the initial magnetic energy to
%  times much greater than have been followed in the simulations, since
%  permanently magnetically trapped particles are the basis for MHD
%  behavior.

%  Because the downstream magnetic turbulence actually is highly
%  intermittent in space, magnetic trapping of a substantial fraction
%  of the particles within the magnetic structures occurs and may be
%  the origin of the discrepancy between the theory and the late time
%  damping observed numerically. We exhibit results from the
%  simulations which demonstrate that magnetic trapping does occur in
%  the downstream medium, since even though the average magnetic energy
%  density is small compared to the thermal energy density, the actual
%  fields in the widely space filaments are strong enough to deflect
%  and trap a large fraction of the particles that encounter the
%  filaments. We speculate that trapping may allow persistence of a
%  small fraction of the initial magnetic energy to times much greater
%  than have been followed in the simulations, since permanently
%  magnetically trapped particles are the basis for MHD behavior.

\end{abstract}


\section{Introduction}

The prompt emission and afterglows of gamma-ray bursts (GRBs) may be
manifestations of ultrarelativistic shock waves, propagating in media
where the large scale upstream magnetization is too weak to affect the
shock structure, and too weak, if simply compressed by the shock, to
provide the magnetization inferred from synchrotron models of the
burst emission (see Piran 2005ab and references therein).  The weakly
magnetized outflows in the rotational equators of rotation powered
pulsars (Coroniti 1990) may also be sites of essentially unmagnetized
shock waves terminating the relativistic winds in a region which
occupies a finite latitude band with respect to the rotational equator
of the underlying neutron star.  Yet the radiation from these systems
has been modeled as being due to synchrotron radiation, which requires
the presence of a magnetic field strong enough to deflect particles
through a substantial fraction of their Larmor orbits. At the very
least, this requires magnetic structures with amplitude, $\delta B$,
whose characteristic dimensions, $R_B$, are comparable to or larger
than $mc^2 /e \delta B$, for all
particle energies inferred to contribute to the observed radiation.

Unmagnetized anisotropic plasmas spontaneously generate small-scale
magnetic fields via the Weibel instability (Weibel 1959).  Shocks have
strong plasma anisotropy in the transition layer separating the
upstream and downstream media, as well as in the foreshock where
downstream particles escape into the upstream, which provides the free
energy for generating magnetic fields with spatial scales on the order
of the plasma skin depth, $c/\omega_p$, where $\omega_p$ is the plasma
frequency.
%These magnetic fields provide the anomalous viscosity, which
%mediates a shock, through particles' elastic scattering from the
%non-propagating (in the shock frame) magnetic fluctuations (Moiseev
%and Sagdeev 1963).  Anisotropic flow energy is converted into
%isotropic themal energy, as measured in the frame of the downstream
%plasma.
Medvedev and Loeb (1999) and Gruzinov and Waxman (1999) argued that the relativistic form of this
instability (Yoon \& Davidson 1987) could macroscopically magnetize
the downsteam plasma where the GRB radiation arises (Larmor radii
small compared to flow scale).  Through numerical and analytic
studies, they and various authors (Silva {\it et al.} 2003, Medvedev
{\it et al.} 2005) have argued that the instability forms filaments of
electric current and $B$ field.  These filaments merge and cause
magnetic energy to cascade from the initial microscopic scale $\sim
c/\omega_p$ to larger scales.  Thus, the filamented plasma becomes
magnetized with a $B$ field hypothesized to survive throughout a large
fraction of the shocked medium.

Such inverse cascades have been observed in 2D and 3D particle-in-cell
simulations in the {\it foreshock} region, the part of the shock
structure where the upstream and downstream media interpenetrate and
the stream filamentation modes grow from noise with most incoming
particles still undeflected from the upstream flow.  
%In the foreshock
%region, the currents appear to self-organize into filaments which
%merge and generate ever larger scale magnetic structures until the
%plasma becomes magnetized.  
Simulations show that current filaments merge and grow in amplitude
until they reach the magnetic trapping limit, where the filament
currents and their magnetic fields become comparable to the Alfven
limit (Davidson {\it et al.} 1972; Kato 2005; also see Milosavljevic,
Nakar, \& Spitkovsky 2006; Milosavljevic \& Nakar 2006a). At that
point the particles' orbits on the filament boundaries become chaotic,
the filaments disorganize, and scattering from the disorganized
magnetic fluctuations halts the streaming of the bulk of the plasma,
isotropizing and thermalizing the flow, all within a layer tens to
hundreds of skin depths thick (Spitkovsky 2005), in accord with Kato
(2005)'s model for shock formation. The magnetic energy is as high as
10-20\% of the bulk plasma flow energy {\it within this scattering
  layer}, where the density jump between upstream and downstream
occurs.

%This picture of merging current filaments
%has grown into a paradigm for Gamma Ray Burst physics, with regular
%invocation of this physical process as the explanation of the magnetic
%fields required in synchrotron models (Piran 2005ab, Katz et al 2006).

These initially small-scale B-fields must survive for tens of
thousands to millions of inverse plasma periods to serve as the source
of the magnetization invoked in synchrotron models of GRB emission
(Gruzinov \& Waxman 1999; Piran 2005ab; Katz, Keshet, \& Waxman 2007). Long-lived fields are also required in Diffusive Fermi Acceleration (DFA)
models used to explain the appearance of the nonthermal particle
spectra observed through their synchrotron and inverse Compton
emission in GRBs, and in pulsar wind nebulae (PWNs) and other sites of
nonthermal photon emission in relativistic flows.  However, present
simulations (Kazimura {\it et al.} 1998; Silva {\it et al.} 2003; Frederiksen  {\it et al.}
2004; Hededal {\it et al.} 2005; Nishikawa {\it et al.}  2003, 2005)
have not followed the flow through the shock transition layer into the
downstream region because they employ periodic boundary conditions,
are too small, or both.  The question of the structure and long-term survival of 
the B-fields has remained open (see for instance, Gruzinov \& Waxman 1999; Gruzinov 2001ab; Medvedev 
{\it  et al.} 2005).

In this paper, we show via linear theory that the phase mixing between
individual particles and organized macroscopic currents implies
rapid decay of the magnetic energy in the downstream medium.  We begin
by briefly describing the essential features of the plasma produced by
the shock via numerical simulations in \S \ref{sec:simulation}.  These
simulations show the downstream magnetic structures are
non-propagating in the frame of the downstream medium, and are intermittent spatially, organized into clumps
and filaments of magnetic field with typical diameter $\sim 10-20$ skin depths, immersed into a highly isotropic plasma.  
%(oriented across the flow direction in the 2D results exhibited here;
%3D results, which form large scale loops and arcs, will be discussed
%elsewhere) 
In this downstream isotropic Maxwellian particle distribution, we
calculate the damping rates from Vlasov linear response theory,
assuming the particles are unmagnetized (Larmor radii $\gg$ magnetic
clump diameter), recovering a result due to Mikhailovskii (1979) in \S
\ref{sec:theory} and Appendix A.  Taking snapshots from the
simulations as initial conditions, we calculate the decay of the
magnetic field as a function of time and position, and compare these
theoretical calculations with simulations.  We find that the theory
does reasonably well in estimating the decay rate of the total
magnetic energy as $t^{-q}$ with $q\sim 1$.  However, it overestimates
the damping rate of shorter wavelength modes.  We speculate in
\S\ref{sec:trap} that this discrepancy may result from the magnetic
trapping of a large fraction of the particles, which suppresses the
disorganizing effect of phase mixing on the currents. We also discuss
the effects of an inverse cascade on the persistence of magnetic
fields, though we find little evidence for its relevance from
numerical simulations.  We summarize our results and draw some
implications for Fermi acceleration and for the magnetization required
for synchrotron emission in models of GRBs and other systems in
\S\ref{sec:discussion}.

%mediate the
%acceleration of a small number of particles to large $\gamma$'s.

%These small scale fields have been invoked as  In order to do so, these 

%In
%this scenario, these particle radiate through synchrotron emission in the
%magnetic fields created at the shock interface.

%For this magnetic energy to survive with little or no decline of the
%magnetic energy for the scale of the post shock flow that gives rise
%to the observed radiation ($10^6 - 10^{10}$ skin depths, Piran 2005)
%requires the now disorganized magnetic structures to continue
%merging to ever larger length scales.  If this is the case, the GRB
%magnetization problem would be solved.  Instead, simulations to date
%have shown that these fields are strongly damped {\it downstream},
%although because of residual electromagnetic noise introduced by the
%finite (although large) number ($10^{8.5} - 10^{10}$)of macroparticles
%employed, the asymptotic state of the magnetic energy remains
%inconclusive.

\section{Simulation Results}\label{sec:simulation}

Spitkovsky (2005) and Spitkovsky and Arons (in prep) describe a series
of 2D and 3D simulations of relativistic shock waves in $e^+e^-$
plasmas, for varying values of the upstream magnetic field $B_1$,
including $B_1 = 0$.  These are Particle-in-Cell (PIC) simulations
(Birdsall and Langdon 1991), using the code {\it TRISTAN-MP},
developed by one of us (AS).  It is a heavily modified descendant of
the publicly available code {\it TRISTAN} (Buneman 1993). We simulate shocks
by injecting cold relativistic plasma particles at one end of a large
domain and allowing the particles to reflect off a fixed conducting
wall at the other end of the box.  In this paper, we present 2D
calculations utilizing boxes as large as 50,000 x 2048 cells with up
to $1.35\times 10^{10}$ particles which allows us to fully resolve
shock formation. Although we track all three velocity and field
components, due to the two-dimensional symmetry only particle
velocities in the plane of the simulation are non-zero for initially
cold flow, and only the out-of-plane component of the magnetic field is
excited by in-plane currents. In-plane electrostatic fields are also
included.

\begin{figure*}
\plotone{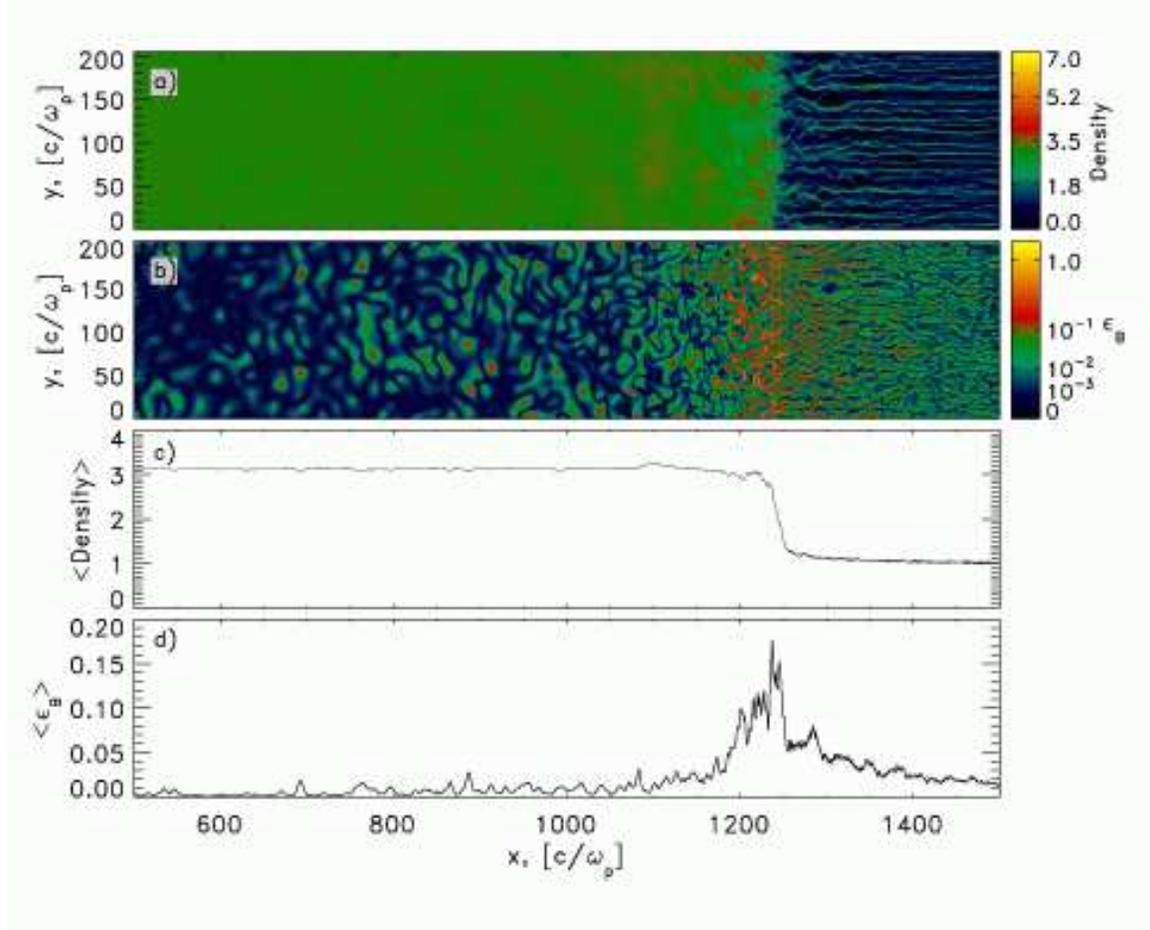}
\caption{Snapshot of a region from a large 2D relativistic shock
  simulation. Incoming $\gamma=15$ flow is moving to the left,
  while the shock moves to the right at $\approx 0.5c$. The postshock
  plasma is stationary in the frame of the simulation.   
a) Density structure in the simulation plane
  showing the plasma density enhancements in the foreshock region that
  steadily grow up to the shock transition region, where the density
  becomes homogeneous. Density is normalized in units of the plasma
  density far upstream. The density
  jump, $\langle n_2 \rangle / \langle n_1 \rangle = 3.13$, shown is
  exactly what is predicted by the hydrodynamic jump conditions for
  $\gamma_1 = 15$ in a 2D gas. b) Magnetic energy, normalized in terms of upstream energy of the
  incoming flow: $\epsilon_B=B^2/4 \pi
  \gamma_1 n_1 m c^2$. 
The upstream magnetic filaments, which can be visualized as
sheets coming out of the page, that are
formed by the Weibel instability reach a peak just before the
shock.  These filaments become clumps of magnetic field, which can be
visualized as cross-sections of loops that are transverse to the page in
the downstream region, where they slowly decay away.  Note that
the B-field, which is organized into upstream filaments
or downstream clumps, always points in or out of the page.
  A power law
  scaling, $\epsilon_B^{1/4}$, was applied to stretch the color table
  to show weak field regions; this is reflected in the colorbar.  c)
  Plasma density averaged in the transverse direction as a function of
  the distance along the flow.  d) Magnetic energy density averaged in
  the transverse direction, as a function of distance along the flow.}
\label{fig:shock}
\end{figure*}

The interaction of the reflected pair plasma with the incoming stream
forms a shock, which propagates toward the plasma injection surface.
In the simulations, one plasma skin depth spans 10 cells based on the upstream parameters,
i.e., $\lambda_{s1} = c/\omega_{p1} = \sqrt{m_\pm c^2 \gamma_1 /4\pi
  e^2 n_{1}}$, where $n_{1} $ is the upstream total density of
electrons and positrons (as measured in the frame of the simulation),
$\omega_{p1}$ is the upstream plasma frequency, and $\gamma_1 m_\pm
c^2$ is the upstream flow energy/particle.  The time step is $\Delta t
= 1/20 \omega_{p1}$.  In the upstream skin depth units, the largest
boxes are then 5000 x 205 $c/\omega_{p1}$. The longest simulation was
evolved for $5300 \omega_{p1}^{-1}$.  We have checked, by using boxes
of increasing transverse dimension, that the periodic boundary
conditions used on the transverse coordinate do not affect the scale
of the magnetic structures formed.  In the results exhibited below, we
use $64$ particles per cell ($32$ per species) in order to suppress
particle noise, which enables us to follow the dynamics of small
amplitude fields. We have performed convergence studies, varying the
number $N$ of particles per cell from $4$ to $64$, confirming that the
noise level decreases as $1/\sqrt{N}$, while the gross qualities of
the shock structure remain the same.  Figure \ref{fig:shock} shows the
snapshots of density and magnetic energy from a typical 2D simulation.\footnote{Note that we have defined the magnetic energy fraction $\epsilon_B\equiv B^2/4 \pi
  \gamma_1 n_1 m c^2$ in terms of
  the {\it upstream} kinetic energy density, instead of the downstream
  thermal energy density as is a common practice in the GRB community
  (c.f., Piran 1999).}  Coordinates are in units of the upstream skin
depth, $c/\omega_{p1}$. In the simulation shown, the upstream flow
moves to the left with $\gamma_1 = 15$.

Our simulations are large enough to permit the complete development of
the shock and show the main features of contemporary collisionless
shock simulations even in reduced dimensionality.  For instance, we
see the factor of $\approx 3.13$ increase in density between the upstream and
the downstream (Fig.  \ref{fig:shock}c), which is the expected compression
factor for this 2D plasma when the plasma properties are measured in
the rest frame of the downstream plasma (Gallant {\it et al.}  1992;
Spitkovsky and Arons, in prep).  Current filaments (geometrically,
these are out-of-plane sheets) show up as the enhancement in plasma
density and magnetic energy density in the foreshock (Fig.
\ref{fig:shock}ab). The scale of the filaments grows towards the shock
through merging.  The shock is located where the density filaments
completely merge and are replaced by a quasi-homogeneous medium. The
subject of the saturation of the Weibel instability will be explored
in greater detail by Spitkovsky and Arons (in prep).

%occurs when
%the
%currents reach a critical strength set by the Alfven current limit (Kato 2005; 
%Davidson {\it et al} 1972),

%fields

%The free particle streaming in the filaments ends when the transverse
%magnetic fields in the filaments grow large enough to deflect
%particles by 90 degrees, a condition known as magnetic trapping, which
%has been long identified as the saturation mechanism for the
%non-relativistic Weibel instability (Davidson {\it et al} 1972).
%Magnetic trapping is equivalent to the Alfven critical current
%condition for saturation when the densities of the two
%counter-streaming beams of particles are the same. 

At the shock transition layer, the filaments disorganize and become
clumps of magnetic energy (in 2D the only appreciable magnetic
component is the out of plane $B_z$).  Note that these magnetic clumps
lose intensity the further downstream they are from the shock (Figure
\ref{fig:shock}b).  The magnetic energy peaks before the density
completes its rise (as we see in comparing c and d in Fig.
\ref{fig:shock}), i.e., the instability saturates at the Alfven
critical current before the shock fully develops.  We also note that
the particle distribution function changes from an anisotropic
free-streaming population to an isotropic (in the downstream rest
frame) thermal population.  In the downstream medium, we find that the
difference between the perpendicular and parallel momentum is
extremely small, $<1\%$.

\begin{figure*}
\plotone{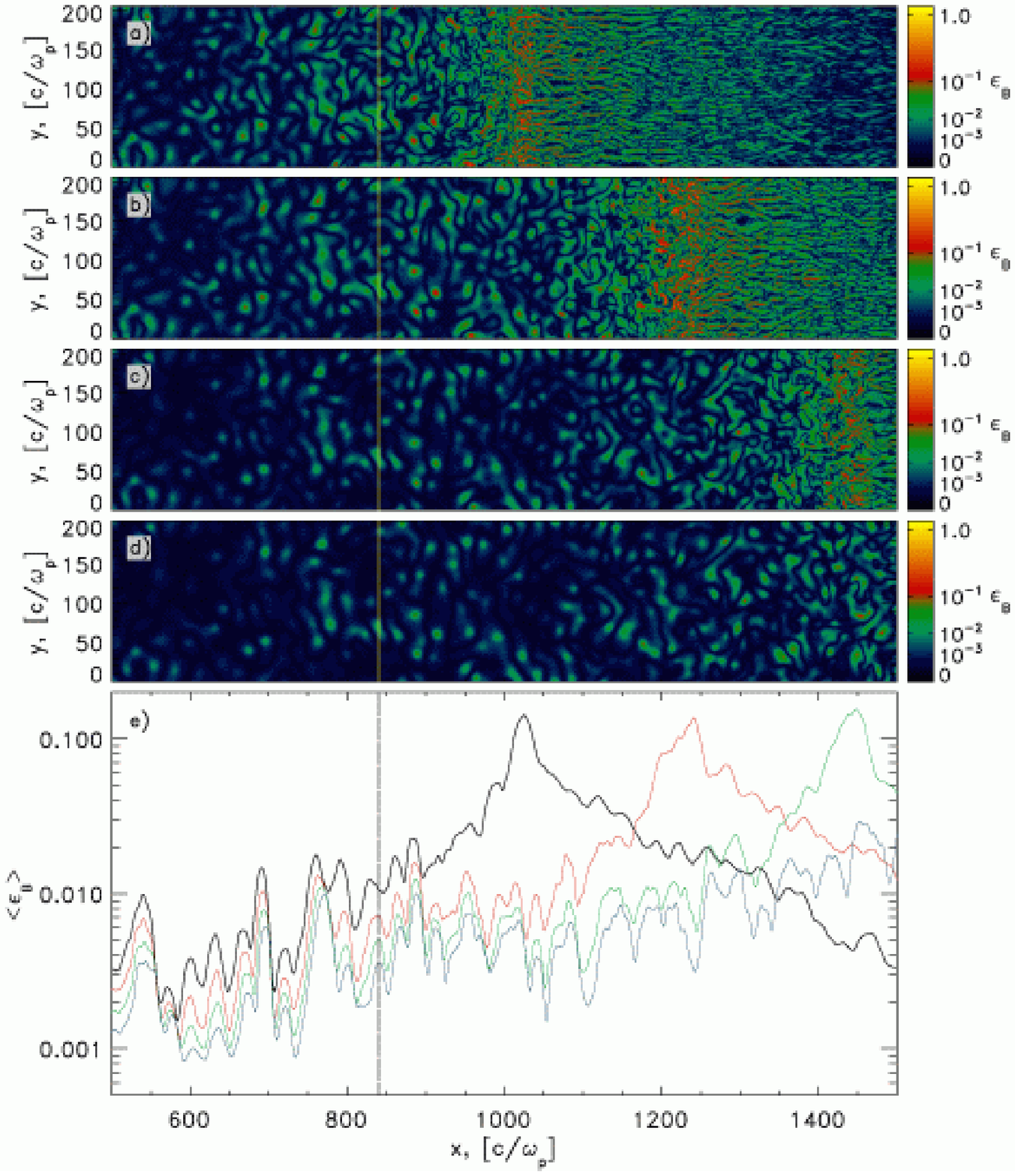}
\caption{ Cross-sectional views of the magnetic energy density for the
  same region as Figure \ref{fig:shock}, but at different times
  separated by 450 $\omega_p^{-1}$ (a-d). The time of Figure
  \ref{fig:shock} corresponds to panel b) in this plot. In panel e) we
  plot the B-field energy, averaged over y-axis as a function of
  distance along the flow direction with one curve for each of the top panels: (a)
  black, (b) red, (c) green, (d) blue.  We also plot a vertical line
  at $x = 840 c/\omega_p$ to define the region over which we perform a
  modal analysis in \S\ref{sec:theory} and Figure
  \ref{fig:fftsingle}.}
\label{fig:shock later}
\end{figure*}

As Figures \ref{fig:shock} b and d suggest, even though locally in the
shock magnetic energy can reach close to equipartition ($epsilon_B \approx 15\%$ when
averaged over the transverse dimension), the magnetic fields decay in
the downstream region of the shock.  To study this in further detail,
we present several cross sections of the plasma at times separated by
450 $\omega_{p1}^{-1}$ in Figure \ref{fig:shock later}. Since our
simulation is performed in the downstream frame (frame of
the reflecting wall), we see the shock propagate through the box at
$\approx 0.5 c$ (value appropriate for 2D relativistic gas).

Downstream from the shock, the magnetic clumps weaken as a function of time
and appear to be non-propagating.  In Figure \ref{fig:shock later}, we see
this in panels a-d and in the black through blue curves in panel e,
which show magnetic energy averaged over the transverse dimension of
the simulation at times from panels a-d.  Although the separation
between prominent clumps that have larger field strength seems to
increase with time, this is due to the faster disappearance of
small-scale clumps, presumably due to decay, while the location or
size of strong clumps does not significantly evolve. There is little
evidence for clump merging far from the shock. In Figure \ref{fig:shock later}
we also plot a vertical line showing the location where we will
decompose the magnetic fields into Fourier modes later in \S3, in
order to study the comparison of the theory of B-field decay with the
numerical experiments.

%There is little evidence
%of merging of the downstream magnetic structures, which now take the
%form of quasi-cylindrical $\theta$ pinches, with B organized in
%filaments directed out of the plane of the figure. 

Finally, we mention that 3D shock simulations also show similar
behavior (including the lack of substantial downstream merging of magnetic
structures) for the shorter times that can be followed with present 3D
simulations.  Topologically, the magnetic clumps in 2D become large
looping structures in 3D. If our 2D plane is thought of as a slice
through a 3D simulation, the 3D loops connect field emerging
from one clump and returning to another. The orientation of 3D loops
would be mostly perpendicular to the original direction of the flow.
The particle distribution function is also an isotropic, relativistic
Maxwellian in the downstream region.

%3D simulations show similar behavior (including lack of substantial
%downstream merging of magnetic structures), for the rather shorter
%times that can be followed. The filaments with straight magnetic
%fields form large looping structures with radii of curvature large
%compared to filament radius, as is illustrated in figure
%\ref{fig:threeDB}.
%
%\begin{figure}[H]
%\plotone{bfield.png}
%\caption{Visualization of the magnetic energy density in a 3D simulation, showing the upstream filaments along the flow to the right and the downstream loops, decaying 
%with distance from the shock front.  }
%\label{fig:threeDB}
%\end{figure}

\section{Downstream Evolution of Magnetic Turbulence}\label{sec:theory}

The simulations summarized in \S\ref{sec:simulation} suggest that
magnetic turbulence decays in the isotropic post-shock plasma.  We now
attempt to understand this theoretically with the goal of finding a model that
allows extrapolation of the magnetic evolution beyond the length of time that can be 
studied in direct shock simulations.  The simulations show that the downstream 
plasma is isotropic and the downstream particle distribution
function is well described by a relativistic Maxwellian. For the
purposes of this section, we {\it assume} the downstream field
amplitudes are so small that magnetic trapping is unimportant for
almost all particles; therefore, their orbits are almost straight
lines.  We will revisit this assumption in \S\ref{sec:trap}.

We begin by deriving the linear plasma response for electromagnetic
fluctuations in an isotropic relativistic plasma with small field
fluctuations (Mikhailovskii 1979).  The Vlasov equation for each species is
\begin{equation}\label{eq:linearize vlasov}
  \frac {\partial f_s}{\partial t} + {\boldsymbol v} \cdot {\boldsymbol \nabla} f_s + 
  \frac {q_s}{m_e} \left({\boldsymbol E} + \frac {{\boldsymbol v}} {c} \times {\boldsymbol B} \right)
  \cdot{\boldsymbol \nabla}_p f_s = 0.
\end{equation}
Here, $s$ is the species label, $+$ for positrons and $-$ for electrons,
with $q_{\pm} = \pm e$.  We linearize this equation with $f_s
\rightarrow f_{0s} + \delta f_s$, ${\boldsymbol B} \rightarrow \delta
{\boldsymbol B}$ and ${\boldsymbol E} \rightarrow \delta {\boldsymbol
  E}$, with the initial conditions downstream of the shock that tell
us that $\delta E$ and $\delta B$ are small in the sense that $(\delta
E^2 + \delta B^2)/8\pi nT \ll 1$.  Then $\delta f_s /f_{0s}$ is also
small.  The linearized Vlasov equation is
\begin{equation}\label{eq:vlasov}
  \frac {\partial\delta f_s}{\partial t} + {\boldsymbol v} \cdot {\boldsymbol \nabla}\delta f_s  
  + \frac {q_s}{m_e} \left(\delta {\boldsymbol E} + \frac {{\boldsymbol v}} {c} \times \delta {\boldsymbol B} \right)\cdot{\boldsymbol \nabla}_p f_{0s} = 0.
 \end{equation}
The plasma couples to the field through the Maxwell equations
\begin{eqnarray}\label{eq:maxwell}
{\boldsymbol \nabla} \times \delta {\boldsymbol E} &=& -\frac 1 c \frac {\partial \delta {\boldsymbol B}} {\partial t},\\
{\boldsymbol \nabla} \times \delta {\boldsymbol B} &=& \frac {4\pi} {c} \delta {\boldsymbol j} + \frac{1}{ c} \frac {\partial \delta {\boldsymbol E}} {\partial t},\label{eq:maxwell2}
\end{eqnarray}
where $\delta {\boldsymbol j} = \sum_s q_s\int {\boldsymbol v} \delta f_s d^3 p$ is the current density.

We orient coordinates so that $\delta {\boldsymbol E}$ is along $x$, the wave vector lies
along $y$, and $\delta {\boldsymbol B}$ is along $z$.  Assuming a wave solution with
amplitudes $\propto \exp\left(-i\omega t + i k y\right)$, equation
(\ref{eq:vlasov}) becomes
\begin{equation}
i \left(kv_y - \omega\right)\delta f_s + \frac {q_s}{m_e} \left(\delta E + 
  \frac {v_y} c \delta B\right)\frac{\partial f_{0s}}{\partial p_x} - 
\frac {q_s}{m_e} \frac {v_x} {c} \delta B \frac {\partial f_{0s}}{\partial p_y} = 0.  
\end{equation}
Using equation (\ref{eq:maxwell}) to eliminate $\delta E$ yields
\begin{equation}\label{eq:deltaf}
  \delta f_s = i \frac {q_s \delta B}{m_ekc} \left(\frac {\partial f_{0s}} {\partial p_x} 
    + \frac {kv_x}{\omega - kv_y} \frac {\partial f_{0s}}{\partial p_y} \right).
\end{equation}
Using (\ref{eq:maxwell2}) yields 
\begin{equation}\label{eq:deltaB}
i\left(\omega^2 - k^2 c^2\right)\delta B = -4\pi kc \delta j.
\end{equation}
Using (\ref{eq:deltaf}) in the definition of $\delta j$ yields
\begin{equation}
\delta j = -i \chi\omega \delta E
\label{eq:j2E}
\end{equation}
where $\chi$ is the plasma susceptibility,
\begin{equation}
  4 \pi \chi    =   \sum_s \frac {\omega_{p,{\rm NR},s}^2}{\omega^2} 
  \int v_x \left(\frac {\partial} {\partial p_x} + \frac {kv_x}{\omega - kv_y} 
    \frac {\partial}{\partial p_y}\right)\frac {f_{0s}}{n_s} d^3p.
\label{eq:susceptibility}
\end{equation}
%and the
%dispersion relation
%\begin{equation}\label{eq:dispersion}
%  \frac {k^2 c^2}{\omega^2} - 1 - \frac {\omega_p^2}{4\pi\omega^2} 
%  \int v_y \left(\frac {d} {dp_y} + \frac {kv_y}{\omega - kv_x} 
%    \frac {d}{dp_x}\right)F_0 d^3p = 0,
%\end{equation} 
Here $n_s$ is the number density of the electrons ($s = -$) or
positrons ($s=+$) in the downstream plasma, $\omega_{p,{\rm NR},s} =
\sqrt{4\pi q_s^2 n_s/m_e}$ is the non-relativistic plasma frequency.
We will assume by charge neutrality that in the downstream plasma
$f_{0+} = f_{0-}$ and $n_+ = n_-$ so that the sum over equal mass
species is trivial.

The dispersion relation for normal modes in the plasma plus electromagnetic 
field is
\begin{equation}\label{eq:general_dispersion}
\frac{k^2c^2}{\omega^2} - (1 + 4 \pi \chi) = 0.
\end{equation}
However, we emphasize that the linear relation between the currents and
the electric field (eq.[\ref{eq:j2E}]) through the susceptibility (\ref{eq:susceptibility}) applies to all 
(small amplitude) fluctuations, not only to normal modes.

%Since $F_0$ is isotropic, we perform the angular integration to find
%\begin{equation}\label{eq:worked out dispersion}
%  \frac {k^2 c^2}{\omega^2} - 1 - \frac {\omega_p^2}{2\omega^2} \int v
%  \left[\left(\frac {\omega} {kv}\right)^2 - \frac{\omega}{kv}\left(1
%  - \left(\frac {\omega} {kv}\right)^2\right)\left\{\frac 1 4
%  \log\left(\frac {1 + \omega/kv}{1 - \omega/kv}\right)^2 - i\frac {\pi}
%  2 \theta\left(\left(\frac {\omega}{kv}\right)^2 - 1\right)\right\}\right]
%  \frac {dF_0}{dp} p^2dp = 0
%\end{equation}

We evaluate equation (\ref{eq:susceptibility}) for distribution
functions that are isotropic in two and three dimensions in Appendix
A.  For a two-dimensional isotropic distribution function, we write
$f_{0s}({\boldsymbol p}) = f(p_{2d})g(p_z)$, where $p_{2d} =
\sqrt{p_x^2 + p_y^2}$ and perform the integral over $p_z$ such that
$\int g(p_z) dp_z = 1$.  For a three-dimensional isotropic
distribution function, we write $f_{0s}({\boldsymbol p}) =
f_{0s}(p_{3d})$, where $p_{3d} = \sqrt{p_x^2 + p_y^2 + p_z^2}$. In the
three-dimensional case, we confirm Mikhailovskii's (1979) result for a
relativistic isotropic plasma.  As we show explicitly in equation
(\ref{eq:worked out suscept}), subluminal waves ($\omega_r < kc$) are
damped, where $\omega_r = \Re(\omega)$.  The basic physics is that of
Landau damping (Stix 1992) or more precisely, phase mixing.
%The standard picture of Landau damping is that there is net power
%transfer between a Fourier mode and a particle when the Doppler
%shifted frequency vanishes, i.e., $\omega_r - {\boldsymbol k} \cdot
%{\boldsymbol v} = \omega_r - k v \cos \phi = 0$.  
For a subluminous wave in an isotropic relativistic plasma with
$\omega_r /k < c$, there always exists some angle $\phi$ between the
wave vector $k$ and a particle's momentum where a particle moving with
speed $c\beta$ is in phase with the field component, i.e.,
$\beta\cos\phi = \omega_r/kc$.

When the phase velocity of the field fluctuations is very small
compared to the thermal speeds of the particles (which includes the
zero wave velocity case, $\Re (\omega) = 0$), Landau damping takes on
the character of simple phase-mixing.  The currents, which support the
field fluctuations, are composed of particles.  These current carrying
particles have random motions which carry them out of the magnetic
structure, which the currents initially support.  As a result, these
currents and their magnetic field fluctuations are disorganized (or
damped) on the transit times of the particles across the field
fluctations.  It should be emphasized that the formal theory includes
both this phase mixing limit of the damping process and the limit in
which the phase 4-speeds $\beta_p /\sqrt{1-\beta_p^2}, \; (\beta_p
\equiv \Re \omega /kc),$ of the fields' Fourier components are high
compared to the particles' random 4-speeds $\beta /\sqrt{1 -\beta^2}$.
The applicable limit depends on whether the wave phase 4-velocity is
large or small compared to the mean 4-velocity of the particles in the
distribution function.  Hammett, Dorland \& Perkins (1992) present a
simple picture of this process in the case of non-relativistic
electrostatic waves.  Arons, Norman \& Max (1977) show how phase
mixing works for electromagnetic waves in a relativistic Maxwellian
plasma.  In any case, the end result is that there is a net power flow
from fields to particles, {\it i.e.} the fields decay, which we will
demonstrate explicitly below.
% Hence as long as population of particle are monotonically decreasing with larger momentum ($dF/dp < 0$) then wave energy is lost.  An inverted population ($dF/dp > 0$) may add energy to the wave, but such a study is beyond the scope of this analysis.  

We now study the action of such a plasma on an arbitary field of
initial B-field pertubations as generated by the Weibel
mediated shock.  We evaluate equation (\ref{eq:susceptibility})
numerically for two and three dimensions setting $\omega_r = 0$,
because of the non-propagating nature of the magnetic clumps, which we
infer from visual inspection of the simulations.  In the limit $k \ll
\omega_p/c$, we find:
\begin{equation}\label{eq:chi_limit}
4\pi\chi \approx \left\{\begin{array}{ll} 
i \frac {\omega_{p}^2}{|k|c\omega} & \textrm{2D} \\
i \frac {\pi} 4 \frac {\omega_{p}^2}{|k|c\omega} & \textrm{3D}
\end{array}\right. . 
\end{equation}
Note that the 2D and 3D results only vary by a numerical factor.
Hence, long wavelength modes have the same qualitative behavior in two
and three dimensions.
%As we show in Appendix A, this simple form recovers the
%exact numerical solution for $\omega_r = 0$.  Hence, we will use
%equation (\ref{eq:chi_limit}) in the rest of the paper.

The plasma susceptibility, which we calculate in Appendix A (see also
eq.[\ref{eq:chi_limit}]), can be utilized to calculate the evolution,
i.e., the damping or growth, of an initial field of fluctuations.  Appendix B
contains this calculation in more detail; the result is
\begin{equation}\label{eq:damping}
\frac {d|\delta B_k|^2}{dt} = -2\gamma_k |\delta B_k|^2,
\end{equation}
where $\gamma_k = \left(kc\right)^2\omega^{-1} \Im\left(4\pi\chi\right)^{-1}
$ (see also eq.[\ref{eq:app_general gamma}]).  The
asymptotic forms of $\chi$ from equation (\ref{eq:chi_limit}) for 3D
and 2D gives (see also eq.[\ref{eq:app_gamma}])
\begin{equation}\label{eq:gamma}
\gamma_{k} = \left\{\begin{array}{ll} \frac {|kc|^3} {\omega_{p}^2} & \textrm{2D} 
\\ \frac {4} {\pi} \frac{|kc|^3} {\omega_{p}^2}& \textrm{3D}
\end{array}\right. .
\end{equation}

\begin{figure}
\plotone{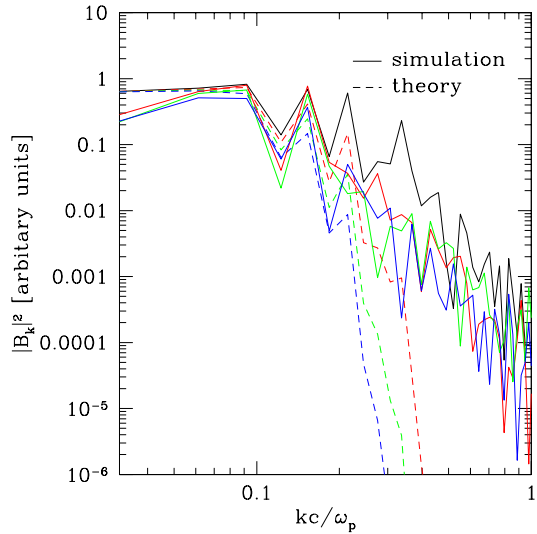}
\caption{ Spectral evolution of magnetic field from the slice at $840 c/\omega_p$ from Fig. \ref{fig:shock later}. Initial field spectrum (black solid line) is plotted after $450\,\omega_p^{-1}$ (red),
$900\,\omega_p^{-1}$ (green), and $1350\,\omega_p^{-1}$ (blue) based on simulation data. Dashed curves represent analytic evolution of the initial field and overpredict decay of short-wavelength structures. 
%Comparison of analytic evolution (dashed line) with the 2D
%simulation (solid lines) at $450\,\omega_p^{-1}$ (red),
%$900\,\omega_p^{-1}$ (green), and $1350\,\omega_p^{-1}$ (blue) of an
%initial field taken from a region behind the shock.  The spectral decomposition of the initial
%field (black solid line) is also shown.
\label{fig:fftsingle} }
\end{figure}

%Linear theory predicts that short wavelengths are
%strongly damped, but long wavelengths persist for longer.  
We now compare the expectations from linear response theory to the
numerical simulations.  We begin by taking the Fourier transform of
$\delta B$ from our 2D numerical simulations from a downstream region
where the shock is fully developed, {\it i.e.}, behind the shock
front at $ x = x_0$,
\begin{equation}\label{eq:fourier transform}
  \delta B(x_0,y) = \frac 1 {\sqrt{2\pi}} \int dk_y \delta B_{k_y}(x_0)
  \exp\left(ik_y y\right)
\end{equation}
We evolve $\delta B_{k_y}$ in accordance to equation
(\ref{eq:damping}) using the asymptotic forms in equation
(\ref{eq:gamma}) and compare our analytically evolved spectra with the
numerical simulation at later times.  In Figure \ref{fig:fftsingle},
we take the initial data from a region at $x_0=840 c/\omega_p$, which
we marked with a line in Figure \ref{fig:shock later}.  To accumulate
sufficient statistics, we average the fields over $14\, c/\omega_p$ in
the flow direction.  We evolve these spectra for 450 (red), 900
(green), and 1350 $\omega_p^{-1}$ (blue) using equation
(\ref{eq:gamma}) and compare this to snapshots taken from our
numerical simulations at these times. Theory and simulation agree at
very low wavenumber ($k_yc/\omega_p \lesssim 0.2$).  However, theory
overpredicts the cutoff in power at larger k.  The discrepancy
suggests that linear theory is insufficient to describe the nature of
downstream magnetic turbulence and that additional physics is needed
(see \S\ref{sec:trap}).

The lack of power in short wavelength modes suggests that the total
B-field is determined by long wavelength modes.  We now use equation
(\ref{eq:damping}) to find a simple decay law for the total B-field:
\begin{equation}
\label{eq:fourier evolution}
  \delta B(x_0,y,t) = \frac 1 {\sqrt{2\pi}} \int dk_y \delta B_{k_y}(x_0)
  \exp\left(ik_y y - \gamma_k t\right),
\end{equation}
where $\delta B_{k_y}(x)$ is defined in equation (\ref{eq:fourier
  transform}).  Inserting equation (\ref{eq:gamma}) into (\ref{eq:fourier evolution}) and
integrating,  with $\delta B_{k_y} =a k_y^p$, where $p$ is the
long wavelength spectral index and $a$ normalizes the amplitude, we find
\begin{eqnarray}
  \delta B(x_0,y,t) &=&  \frac 1 {\sqrt{2\pi}} a \int_{k_0}^\infty dk_y k_y^p
  \exp\left(-\alpha \omega_{p}t \left(\frac {k_yc} 
      {\omega_{p}}\right)^3\right) \\ 
  &\approx& \frac{1}{3 \sqrt{2\pi}} a \left( \frac{\omega_{p}}{c} \right)^{p+1}
     \Gamma\left(\frac {p+1}{3} \right)  
  \left(\frac{1}{ \alpha \omega_{p} t } \right)^{(p+1)/3}
\label{eq:decay-power}
\end{eqnarray}
where we have taken $y=0$ without loss of generality, $\alpha = 4/\pi$
in 3D and $\alpha = 1$ (2D), $k_0 $ is small such that $\omega_p t
(k_0 c/\omega_p)^3 \approx 0$, and $\Gamma$ is the gamma function.
Note that $\omega_p$ is for the downstream plasma frequency.  However, this
detail is irrelevant in terms of determining the power law as a
function of $t$.
% - in the simulations, where time is
%measured in units of $\omega_{p1}^{-1}$ and length in units of
%$c/\omega_{p1}$, the transverse size is $200 c/\omega_{p1}$ in the 2D
%models.  
Setting $k_0$ to be small is a safe approximation in the simulations,
and of course is excellent in much larger astrophysical systems.
Thus if the initial spatial spectrum is a power law in wave number
$|\delta B_k|^2 \propto k^{2p}$, then the theory predicts $\delta B^2
\propto t^{-2(p+1)/3}$.

\begin{figure}
\plotone{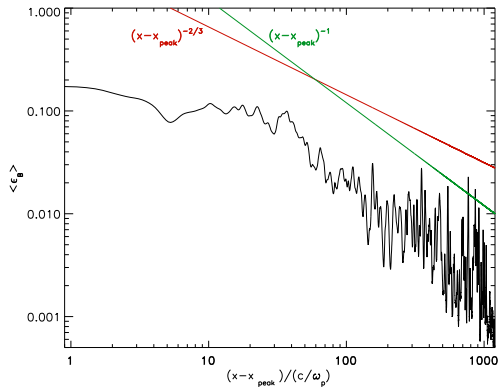}
\caption{Magnetic energy density (in units of upstream kinetic energy) as a function of position downstream of the shock. A broken power law proportional $(x-x_{\rm
    peak})^{-2/3}$ fits well at early times, but a $(x-x_{\rm
    peak})^{-1}$ power law fits better at later times.  }
\label{fig:eb}
\end{figure}

%Magnetic field energy appears initially to follow this trend in our
%numerical simulations as shown in .  
We study the region immediately following the peak of magnetic energy
in the shock front (where it reaches $\delta B^2 = B_{\rm max}^2$) at
$x_{\rm peak}$ and plot its value as a function of position in the
postshock region.  For a shock moving at constant velocity, we have
$x_{\rm peak} - x \propto t$.  Hence $\delta B^2 \propto (x_{\rm peak}
- x)^{-2(p+1)/3}$.  Our numerical simulations are extremely suggestive
that the magnetic energy density follows the $p = 0, \; \epsilon_B =
\delta B^2/8\pi \propto t^{-2/3}$ decay expected for an initially flat
magnetic spectrum at early times, then steepens to a $t^{-1}$ decay at
later times as shown in Figure \ref{fig:eb}, until the noise finally
swamps the signal.\footnote{Gruzinov (2001a) performed shock
  simulations (initiated by a collision between two $e^\pm$ plasmas),
  which show decay of the magnetic energy averaged over the dimension
  across the flow similar to the early phases of the decay we report
  here.}  
This $t^{-1}$ decay rate would imply an initial spatial
index of $p=1/2$ at low wavenumber, though our analysis of additional
simulations with a large transverse spatial scale suggest $p=0$. The 
difference in the index of the decay law expected from the theory and measured from 
the simulations may be due to magnetic trapping (see \S\ref{sec:trap}).
Gruzinov (2001b) offered an alternative
explanation of $t^{-1}$ decay of the magnetic energy density observed
in 2D simulations of Weibel instability in counterpropagating pair
plasmas.  In that simulation the beams were moving perpendicular to
the simulation plane\footnote{Being orthogonal to the direction of
  motion these simulations did not form a shock and retained some
  counterstreaming at late times. In contrast, our simulations lose
  the counterstreaming once the shock forms.}.  In order to explain
the field decay, Gruzinov (2001b) had to assume that the magnetic
structures increase their size at the Alfven velocity. In our shock
simulations, the downstream magnetic structures do not significantly
expand (\S2), hence the similarity of decay law between the two
simulations is likely a coincidence. The picture of filament expansion
and merging of Gruzinov (2001b) is likely valid very close to the
shock, but is not observed in the longer evolution of the downstream
plasma.
%However, his argument assumed that the expanding
%magnetic structures, which he observed in his colliding beam
%simulations,\footnote{Gruzinov (2001b) simulates the evolution of
%  magnetic field during Weibel instability driven by collision of two
%  infinite plasma streams.  The plane of the 2D simulation is
%  perpendicular to the direction of motion of the streams. As such,
%  this simulation does not form a shock.  In contrast, our simulation
%  is in the plane along the motion of the streams, and allows density
%  compression. }  increase their size at the Alfven velocity.  As we
%have pointed out in \S2, these magnetic structures neither propragate
%nor significantly expand.  Hence, Gruzinov's reasoning is not
%applicable to our results.
%We
%discuss Gruzinov's (2001) alternative explanation of $t^{-1} $ decay
%of the magnetic energy density, based on conservation of the vector
%potential (which is a much more severe requirement than conservation
%of canonical momentum), in Appendix \ref{sec:canonical momentum}.

%However, the 2D spectra are rather noisy, since they are 1D spectra over a single
%transverse dimension.  Spectra from 3D simulations are rather better behaved,
%since they are taken over a 2D y-z plane, as shown in Figure \ref{fig:fftpeak}
%These spectra  reveal an interesting departure of the theory from the experiments,
%as shown in Figure \ref{fig:fftpeak}
%\begin{figure}[H]
%  \epsscale{1.0} \plotone{manyBkwithfitbw.png}
%  \caption{Decay of the magnetic fluctuation power spectra  from a 3D simulation
%  run with $512 \times 512 \times 6000$ cells, and 4 particles per cell initially.
%   The four curves show the spectra with time increasing from top to bottom. Each curve 
%   is separated by $\Delta t = 182/\omega_{p1} $.  The fitted straight lines are
%   $| \delta B_k|^2 \propto \exp[- g \omega_{p1}t (kc/\omega_{p1}) ]$ - the decay
%   is better fit with a damping rate linear rather than cubic in $k$. In addition, these
%   3D spectra show substantial magnetic merging at early times, shown by the move of
%   the magnetic energy peak to smaller wave number. This merging dies away as the
%   decaying magnetic clumps become more spatially isolated. }
%\label{fig:fftpeak}
%\end{figure}
\begin{figure*}
\plotone{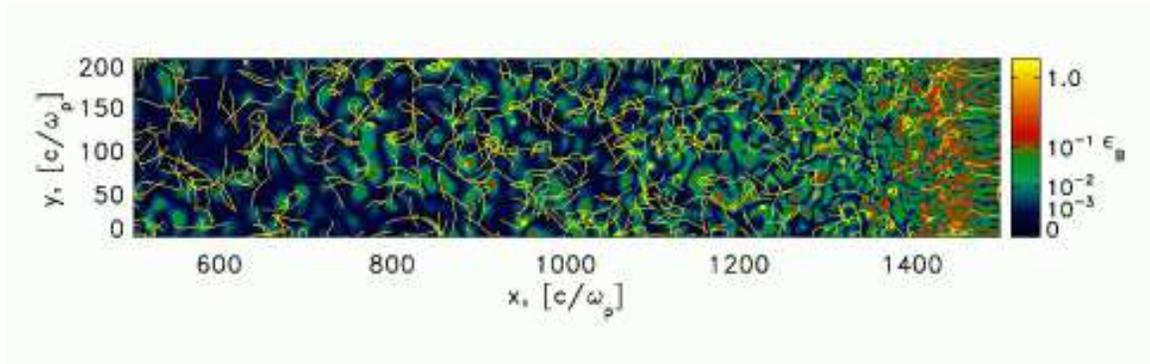}
\caption{Segments of particle orbits from the high resolution 2D simulation overplotted on a snapshot of magnetic energy downstream of the shock. Typical sizes of magnetic clumps are
  10-20 $c/\omega_p$. The shock is near the right boundary of this slice, moving to the right. The large angle deflections in particle orbits suggest the Larmor
  radii of many particles in this slice %$\sim 300 c/\omega_{p1}$
are of the same order as the sizes of the magnetic clumps. 
\label{fig:trapped}}
\end{figure*}

\section{Magnetic Trapping}\label{sec:trap}

Our simulations and theory suggest power law $t^{-q}$ with
$q \sim 1$ temporal decay of the total magnetic energy density in
rough agreement with each other.  However, linear theory and numerical
simulations disagree on the decay rate for short wavelengths, which
hinders a determination of the ultimate fate of the fields on times
longer than several thousand plasma periods. This discrepancy may
arise from the nonlinear effects of magnetic trapping.  Particles in
the magnetic fields do not follow straight line trajectories that are
weakly perturbed, but are partially trapped and strongly deflected.
We illustrate this point from following test particles' orbits in
numerical simulations in Figure \ref{fig:trapped}.  The test particles
suffer large deflections from straight-line orbits as they encounter
magnetic clumps.  Therefore, the Larmor radii of many of the test
particles are of the same order of the sizes of these clumps or
smaller.  Indeed closer inspection of some of the test particle orbits
that are embedded inside the clumps suggests that they are completely
trapped.

%To hammer this point further, we plot the distribution of Larmor radii
%in the whole downstream region in Figure \ref{fig:larmor}.  Since the
%rough size scale of these magnetic clumps are 50-100 cells, a
%substantial number of particles are trapped in these clumps.
%\begin{figure}
%\plotone{LarmorRadiiDistrLinear.png}
%\caption{Distribution of downstream particles' formal Larmor radii
%  $m_\pm c^2 \gamma/eB $. Since the scale of the magnetic structure is
%  typically 50-100 cells, a substantial fraction of particles are likely trapped. 
%  \label{fig:larmor}}
%\end{figure}

These strong departures from weakly perturbed particle dynamics may be
the cause of the decreased damping at large wavenumber found in the
simulations, compared to the predictions of unmagnetized plasma theory.
Magnetic trapping may also modify the decay of magnetic fields at
small wavenumber, leading perhaps to a different decay law instead of 
expected $t^{2/3}$ decay law that we find from our simple linear theory 
(eq.[\ref{eq:decay-power}] with $p=0$).
The temporary (and permanent, for some particles)
binding of particles to the spatially intermittent magnetic fields
reduces the effect of rapid phase mixing central to the unmagnetized
plasma damping theory.  Magnetic trapping already plays a critical
role in the saturation of the initial Weibel instability at the shock
transition region (Kato 2005; Davidson {\it et al.}  1972). Our
analysis suggests that trapping also plays an important role
downstream even though the {\it average} magnetic amplitudes are
greatly reduced from the shock transition region - the essential point
is that within the isolated filaments, the magnetic pressure is not
small.  The problem of the damping of isolated magnetic structures in
a plasma with partial magnetization illustrated in Figure
\ref{fig:trapped} will be the subject of a separate investigation.

\section{Discussion}\label{sec:discussion}

We have studied the downstream evolution of magnetic turbulence in the
context of a collisionless $e^+e^-$ shock both analytically and
numerically.  Our large scale 2D simulations show the formation of
filaments in the foreshock region which merge and grow
until they reach the shock transition region.  Past the shock
transition region, these filaments break up into magnetic clumps in a
quasi-homogenous medium where the background particle distribution
function is an isotropic Maxwellian.  In such a background, we showed
that magnetic energy will decay like $t^{-q}$ with $q\sim
1$ due to untrapped particle phase mixing, which is broadly consistent
with our numerical simulations.  In detail, the theoretical decay
rates depend more strongly on wavelength than is seen in the numerical
experiments.  We suggested that magnetic trapping may play an
important role in resolving this discrepancy.  Trapping can lead to
MHD-like behavior and possible long time persistence of some of the
magnetic energy in spatially intermittent, more strongly magnetized
subregions. This effect is a subject of further study.

% In addition, a
%sufficiently fast inverse cascade may reduce magnetic decay.  However,
%our simulations show no evidence for such a mechanism in the
%downstream medium.

Magnetic field decay may be alleviated via an inverse cascade from
small scales to large scales.  An inverse cascade via current filament
merging operates strongly in the {\it foreshock} region (Silva {\it et
  al.}  2003; Spitkovsky 2005).  However, it is unclear if this
picture can be applied to the downstream region.  The magnetic
filaments so prevalent in the upstream region are gone and are
replaced by isolated magnetic clumps (loops), a result common to both 2D and
3D simulations.  Filament merging does occur, but is confined to the
foreshock, where the filaments exist.  We see little evidence of
merging magnetic clumps.  This is reinforced by Katz {\it et al.} (2007), who
suggest via self-similar arguments that the inverse cascade does not
operate downstream of the shock.  Hence, the magnetic energy should damp
away in the downstream region.

We have not studied in detail how these magnetic clumps are confined
in the downstream plasma.  However, Fujita et al. (2006) have found
these same magnetic clumps in simulation of the non-relativistic
Weibel instability that is appropriate for galaxy clusters.  In this
case, they argue that pressure of the external medium confines these
magnetic clumps.  The strong perturbation from straight line motion,
which we studied in \S\ref{sec:trap} for test particles, suggest that
the magnetic clumps in our simulation are also sensitive to momentum
transfer between particles and itself.  That is, the magnetic clumps
in our simulations are also confined by external pressure.

Whatever the final fate of the spatially intermittent fields, our
study of the nature of $e^+e^-$ shocks suggests that the belief in the
persistence of Weibel generated magnetic fields with strengths
comparable to those appearing within the shock transition is overly
optimistic.  Magnetic field energies tend to decline rapidly after
about a few hundred plasma skin depths. Only the very long term
evolution is still open to some question, i.e., does $\langle
\epsilon_B \rangle$ settle at some value smaller than $10^{-3}$, or
does the spatially intermittent magnetic field decline to zero?

The rapid decay suggested by our analysis puts severe constraints on
the synchrotron emission mechanisms for GRBs.  The width of the
emitting region is $\sim 10^9 c/\omega_p$ (Piran 2005b), which is much
larger than the region over which we expect magnetic fields to
persist.  However, decaying magnetic fields may not be inconsistent
with GRB observations.  Pe'er \& Zhang (2006) suggest that the prompt
emission from the internal shock may be more consistent with a
decaying magnetic field component rather than a persistent field
component.  A field that persists over a scale of $10^4-10^5$ plasma
skin depths fits the spectra better at low-energies than a
field that persists over the entire thickness of the shell ($10^9$
plasma skin depths).  Small magnetized regions may also be important
in the context of the afterglow (Rossi \& Rees 2003).

In young pulsar wind nebulae, post
shock magnetic fields averaged over the whole latitudinal extent of
the observed emission tori have energy densities of a few
percent of the post shock plasma energy density.  It is possible that
weaker magnetic fields exist near the midplane of the equatorial flow, a
region of particular interest to the conversion of flow energy into
the observed nonthermally emitting spectra of $e^\pm$.  If so, the
Weibel mediated shock dynamics studied here may be of relevance to
these systems' behavior.

It is also possible that weak systematic upstream magnetic fields are
of essential importance, and that shocks in completely unmagnetized
plasmas are an oversimplification.  Suprathermal particle generated at
the relativistic shock front may alter the basic physics of the
collisionless shock, if a mean field is present.  Milosavljevic and
Nakar (2006b) argue accelerated particles streaming into the {\it
  upstream} magnetized medium can drive long wavelength, magnetized
turbulence with $\delta B /B \gg1$, which might persist into the
downstream and provide the magnetization required in phenomenological
models of GRB and PWN emission.  Then the shock mediated by Weibel
turbulence becomes a subshock within a much larger extended structure,
responsible only for thermalizing the bulk of the flow and injecting
the particles that are Fermi accelerated in the turbulence generated
by high energy particle streaming.  This is a relativistic version of
Bell's (1978) (also see Bell 2004, 2005) picture of particle
acceleration in non-relativistic shocks.

Finally, we briefly mention the extension of this theory to
electron-ion plasmas.  Let us presume initially that the ions and
electrons are isotropic but remain decoupled, i.e., a two-temperature
relativistic plasma. Equation (\ref{eq:chi_limit}) becomes
\begin{equation}\label{eq:i-e chi_limit}
\chi \approx i\frac {\pi} 4 \frac {\omega_{p,i}^2 + \omega_{p,e}^2}{|k|c\omega},
\end{equation}
where $\omega_{p, i}^2 = 4\pi n_i e^2/\gamma_i m_i$ is the plasma
frequency of the ions, $\gamma_i$ is the Lorentz factor associated
with the ion temperature, $\omega_{p, e}^2 = 4\pi n_e e^2/\gamma_e
m_e$ is the plasma frequency of the electrons, and $\gamma_e$ is the
Lorentz factor associated with the electron temperature.  Depending on
the relative values of the electron temperature and ion temperature,
one term may dominate.  However, initial large-scale simulations of
ion-electron collisionless shocks suggest that both reach roughly
equipartition with each other (Spitkovsky 2007), thereby reproducing
the physics of the $e^{\pm}$ shock.  In this case, the relativistic
electrons and ions contribute equally to the decay rate because
thermal equipartition prevails, i.e., $m_e \gamma_e = m_i
\gamma_i$. Thus, the electron-ion plasma has the same dynamics as the
electron-positron plasma.

\acknowledgements

We thank S. Cowley, D. Kocelski, M. Milosalavjic, A. Pe'er and E.
Quataert for useful discussions. P.C. and J.A. thank the Institute for
Advanced Study for its hospitality; P.C. also thanks the Canadian
Institute for Theoretical Astrophysics for similar hospitality.  P.C.
is supported by the Miller Institute for Basic Research. J.A. has
benefited from the support of NSF grant AST-0507813, NASA grant
NNG06GI08G, and DOE grant DE-FC02-06ER41453, all at UC Berkeley; by
the Department of Energy contract to the Stanford Linear Accelerator
Center no. DE-AC3-76SF00515; and by the taxpayers of California.  A.S. is pleased to acknowledge that the simulations reported on
in this paper were substantially performed at the TIGRESS high
performance computer
center at Princeton University which is jointly supported by the Princeton
Institute for Computational Science and Engineering and the Princeton
University Office of
Information Technology.
\appendix 
\section{Susceptibility in Two and Three Dimensions}\label{sec:susceptibility}

In this section, we solve the susceptibility (eq.[\ref{eq:susceptibility}]) for two and three dimensional plasma.  The 3D case
(\S\ref{sec:3dresult}) has been previously solved by Mikhailovskii
(1979) and we reproduce his result here for completeness.  In
addition, we present the 2D case (\S\ref{sec:2dresult}), which
allows for a comparison to the two-dimensional simulations reported in
this paper.

\subsection{Three Dimensions}\label{sec:3dresult}

After summing over the electrons and positrons, equation (\ref{eq:susceptibility}) is
\begin{equation}\label{eq:susceptibility 3d}
  4 \pi \chi    =   \frac {\omega_{p,{\rm NR}}^2}{\omega^2} 
  \int v_x \left(\frac {d} {dp_x} + \frac {kv_x}{\omega - kv_y} 
    \frac {d}{dp_y}\right)\frac {f_0}{n} d^3p,
\end{equation}
where $n$ is the number density of electrons and positions.  Since
$f_0$ is isotropic and therefore independent of angle, we orient the
spherical integral in a {\it non-standard manner} so that the pole
points along the k-vector, i.e., the y-axis.  We find
$\boldsymbol{\hat{p}}\cdot \boldsymbol{\hat{y}} = \cos\theta$ and
$\boldsymbol{\hat{p}}\cdot \boldsymbol{\hat{x}} = \sin\theta\sin\phi$.
So equation (\ref{eq:susceptibility 3d}) becomes
\begin{equation}
  4 \pi \chi    =   \frac {\omega_{p,{\rm NR}}^2}{\omega^2n} 
  \int v\sin^2\theta\sin^2\phi \left(1 + \frac {\cos\theta}{\omega/kv - \cos\theta}\right)\frac {df_0} {dp} p^2dpd\Omega,
\end{equation}
where $d\Omega = \sin\theta d\theta d\phi$.  Performing the integral
over $\phi$ and making the substitution $\zeta = \cos\theta$, we find
\begin{equation}
  4 \pi \chi   =   \frac {\pi\omega_{p,{\rm NR}}^2}{\omega^2n} 
  \int v\left(1-\zeta^2\right) \left(1 + \frac {\zeta}{\omega/kv - \zeta}\right)\frac {df_0} {dp} p^2dpd\zeta.
\end{equation}
Integrating over $\zeta$ from -1 to 1, we find
\begin{eqnarray}
  4 \pi \chi  &  = & \frac {2\pi\omega_{p,{\rm NR}}^2}{\omega^2n}  \int  p^2dp \frac {df_0}{dp}  v{\Bigg \{}  \left(\frac
  {\omega} {kv}\right)^2 - \frac{\omega}{kv}\left[1 - \left(\frac{\omega} {kv}\right)^2\right] 
  \nonumber \\
 &  \hspace*{0cm}  & \hspace{1cm} \times \left[ \frac{1}{ 2} \log \left(-\frac
  {1 + \omega/kv}{1 - \omega/kv}\right)\right] {\Bigg \}}.
   \label{eq:worked out suscept1}
\end{eqnarray}
Note that the logarithmic function will give an imaginary part when
$\omega_r/kc\beta < 1$. This implies that waves whose phase velocity,
$\omega_r/k$, is small compared to the thermal speed of the background
particles, $c\beta$, will be damped (also see \S\ref{sec:theory}). We
pull this imaginary component out of the equation, which makes this damping
more explicit, to find:
\begin{eqnarray}
  4 \pi \chi  &  = & \frac {2\pi\omega_{p,{\rm NR}}^2}{\omega^2n}  \int  p^2dp \frac {df_0}{dp}  v{\Bigg \{}  \left(\frac
  {\omega} {kv}\right)^2 - \frac{\omega}{kv}\left[1 - \left(\frac{\omega} {kv}\right)^2\right] 
  \nonumber \\
 &  \hspace*{0cm}  & \hspace{1cm} \times \left[ \frac{1}{ 4} \log \left(\frac
  {1 + \omega/kv}{1 - \omega/kv}\right)^2 - 
   i\frac {\pi}{2}
  \Theta\left(k^2 v^2 - \omega_r^2 \right)\right] {\Bigg \}}
  , \label{eq:worked out suscept}
\end{eqnarray}
where $\Theta$ is the unit step function (i.e., $\Theta(z) = 1$ for
$\Re(z) \geq 0$ and $\Theta(z) = 0$ for $\Re(z) < 0$).
Equation (\ref{eq:worked out suscept}) is precisely Mikhailovskii
(1979)'s result.  
We now apply a three dimensional relativistic Maxwellian $f_0 \propto
\exp\left(-{E}/kT\right)$, which is appropriately normalized, $\int
d^3p f_0 = n$, and solve equation (\ref{eq:worked out suscept})
numerically.  For the ultrarelativistic case $v\approx c$, we find a
simple form in the limit $\omega_r \ll kc$:
\begin{equation}\label{eq:3dchi}
4\pi\chi \approx i\frac {\pi} 4 \frac {\omega_{p}^2}{|k|c\omega}. 
\end{equation}

\subsection{Two Dimensions}\label{sec:2dresult}

Starting from equation (\ref{eq:susceptibility 3d}), we assume $f_0 =
f(p_{2d})g(p_z)$, where $p_{2d} = \sqrt{p_x^2 + p_y^2}$ and perform
the integral over $p_z$.  The resulting two dimensional analogue of
equation (\ref{eq:susceptibility 3d}) is 
\begin{equation}\label{eq:susceptibility 2d}
  4 \pi \chi =  \frac {\omega_{p,{\rm NR}}^2}{\omega^2} 
  \int v_x \left(\frac {d} {dp_x} + \frac {kv_x}{\omega - kv_y} 
    \frac {d}{dp_y}\right)\frac {f} n d^2p,
\end{equation}
where we have dropped the subscript ``$2d$'' from $p$.  Defining $p_x =
p\cos\theta$, $p_y = p\sin\theta$, and similarly for $v_x$ and $v_y$,
we find:
\begin{equation}
  4\pi\chi= \frac {\omega_{p,{\rm NR}}^2}{\omega^2 n}
  \int pdp \int_0^{2\pi} d\theta v\cos^2\theta \left(1 + \frac
  {\sin\theta}{(\omega/kv) - \sin\theta} \right)\frac{df}{dp}.
\end{equation}
We may transform the $\theta$ integral from $0$ to $2\pi$ to a contour
integral over the unit circle by making the appropriate substitutions
(Carrier, Krook, \& Pearson 1983)
\begin{eqnarray}
\cos\theta &\rightarrow& \frac 1 2\left(z + z^{-1}\right), 
\sin\theta \rightarrow \frac 1 {2i}\left(z - z^{-1}\right) \\
d\theta &\rightarrow& \frac {dz} {iz}, 
\int_0^{2\pi} \rightarrow \int_{\Gamma},
\end{eqnarray}
where $\Gamma$ is the unit circle.  After a bit of algebra, we find
\begin{equation}
  4\pi\chi= \frac {\omega_{p,{\rm NR}}^2}{\omega^2n}
  \int v \frac{df}{dp} pdp \int_{\Gamma} \frac {-i} 4 dz 
  \left(z^2 + 2 + z^{-2}\right)\frac
  {(2i\omega/kv)}{(2i\omega/kv)z - z^2 + 1}.
\end{equation}
The singular points in this equation are $z_{\pm} = (i\omega/kv) \mp
\sqrt{1 -(\omega/kv)^2}$.  We perform the contour integral by noting
that only the $z_-$ root contributes for $(\omega/kv)^2 < 1$:
\begin{equation}\label{eq:2-d susceptibility}
  4\pi\chi = -i\frac {2\pi\omega_{p,{\rm NR}}^2}{\omega k n}
  \int \sqrt{1 - \left(\frac{\omega}{kv}\right)^2} \frac{df}{dp} pdp.
\end{equation}
We apply a two dimensional relativistic Maxwellian $f \propto
\exp\left(-{E}/kT\right)$, where $f$ is appropriately normalized, i.e.
$\int f d^2p = n$.  Expanding to lowest order in $\omega/kv$, we find:
\begin{equation}\label{eq:2-d result}
4\pi\chi \approx i\frac {\omega_{p}^2}{|k|c\omega}.
\end{equation}

\section{Decay of an Initial Field of Fluctuations}\label{sec:fluctuation dissipation theorem}

The plasma susceptibilities (eq.[\ref{eq:chi_limit}]) define the
linear response of the plasma.  These susceptibilities are complex and
hence the electric permittivity $\epsilon = 1 + 4\pi\chi$ is also
complex.  The implications of Poynting's theorem for the propagation 
of small amplitude waves and fluctuations in such a medium have been
described in many texts and monographs (e.g., Bekefi 1966, Melrose and
McPhedran 1991, Stix 1992).  For completeness, we give a brief discussion
of the theory behind expression (\ref{eq:damping}).

We begin with Poynting's theorem:
\begin{equation}
\frac{\partial}{\partial t} \left(\frac{\delta B^2 + \delta E^2}{8\pi}
     \right) + {\boldsymbol \nabla}\cdot{\boldsymbol S} = - {\delta
     \boldsymbol j} \cdot \delta {\boldsymbol E},
\end{equation}
where ${\boldsymbol S} = (c/4\pi){\boldsymbol E}\times{\boldsymbol B}$
is the Poynting vector.  We assume field energy is distributed
uniformly in our uniform plasma, so ${\boldsymbol
\nabla}\cdot{\boldsymbol S} = 0$, i.e., there is no transport of
energy spatially. In addition, for nonpropagating Weibel modes, $\delta
E \ll \delta B$.  Thus, we find a simplified expression:
\begin{equation}
\frac{\partial}{\partial t} \left(\frac{\delta B^2}{8\pi} \right) = -
     {\delta \boldsymbol j} \cdot \delta {\boldsymbol E}.
\end{equation}
Writing the fields with truncated amplitudes
\begin{equation}  \delta B_{TV} (  {\boldsymbol r}, t)  = \left\{
     \begin{array}{rl}
    \delta B, \; t \in (-T/2, T/2), \;  |{\boldsymbol r}|   \in  V, \\
    0, \; t \not\in (-T/2, T/2), \;  |{\boldsymbol r}|  \not\in  V,   
   \end{array} 
   \right.  
\end{equation}
with $\delta B_{TV} = 0$ when the coordinates are outside of the
volume, $V$, and time is outside of the interval $(-T/2, T/2)$. Thus,
we can define space-time averages of the fields while still expressing
them in terms of convergent Fourier transforms.  We write the Fourier
transforms as
\begin{equation}
 \delta j_{k\omega} = \int d^3k \int_{-\infty}^{\infty}d\omega\,\delta j_{TV} 
\exp\left(i{\boldsymbol k}\cdot{\boldsymbol r} - i\omega t\right).
\end{equation}
We express the field energy density and the $\delta j \cdot \delta E$
work in terms of the Fourier amplitudes and then average over time and
space. Taking $T$ and $V$ to $\infty$, we integrate over the resulting
$\delta$-functions to find
 \begin{equation}
\frac{\partial}{\partial t} \left \langle \frac{\delta B^2}{8\pi} \right \rangle  = 
        \int d^3k\int d\omega\frac 1 2 
       \langle \delta j_{k\omega } \delta E_{k\omega}^* + c.c. \rangle,
\end{equation}
where $\langle \rangle$ represents the space time average over the
fluctuation wavelengths and variability times and $c.c.$ is the
complex conjugate.  We apply the same Fourier transform to $\delta B$
and apply the same average over time and space.  We use the linear
response (eq.[\ref{eq:j2E}]) $\delta E_{k\omega} = (i / \chi \omega)
\delta j_{k\omega}$ to find
\begin{equation}
\frac 1 {8\pi}\frac{\partial \langle |\delta B_{k\omega}|^2 \rangle}{\partial t}  
= -\Im\left(\omega \chi\right)^{-1} \langle |\delta j_{k\omega} |^2 \rangle. 
\end{equation}
Now using $\delta j_{k\omega} = ikc \delta B_{k\omega}/4\pi$,
we find
\begin{equation}\label{eq:app_damping}
\frac{\partial  \langle |\delta B_{k\omega}|^2 \rangle }{\partial t} 
  = -2 \gamma_{k\omega} \langle |\delta B_{k\omega} |^2 \rangle,
\end{equation}
where $\gamma_{k\omega}$ is the damping rate or
\begin{equation}\label{eq:app_general gamma}
  \gamma_{k\omega} = \frac {\left(kc\right)^2}{\omega} \Im\left( 4\pi\chi\right)^{-1}.    
\end{equation}
Equation (\ref{eq:app_damping}) is a specific form of the
fluctuation-dissipation theorem (Thompson and Hubbard 1960) from which
the same result can be derived.
 
Equations (\ref{eq:app_damping}) and (\ref{eq:app_general gamma})
define the evolution of magnetic energy in terms of the linear
response.  For notational simplicity we reduce $k\omega \rightarrow k$
in the rest of the text.  We obtain simple forms for $\gamma_k$, using
the asymptotic forms of $4\pi\chi$ from equation (\ref{eq:chi_limit}) for
3D and 2D. We find for $kc \ll \omega_p$:
\begin{equation}\label{eq:app_gamma}
\gamma_{k} = \left\{\begin{array}{ll} \frac {|kc|^3} {\omega_{p}^2} & \textrm{2D} 
\\ \frac {4} {\pi} \frac{|kc|^3} {\omega_{p}^2}& \textrm{3D}
\end{array}\right. .
\end{equation}
The cubic dependence on wavenumber in equation (\ref{eq:app_gamma})
suggests short wavelength modes are very strongly damped, while longer
wavelength modes can survive much longer.  
%In addition, the 2D and 3D
%damping rates are the same up to a overall multiplicative constant of
%order unity.

\begin{figure}[H]
\plotone{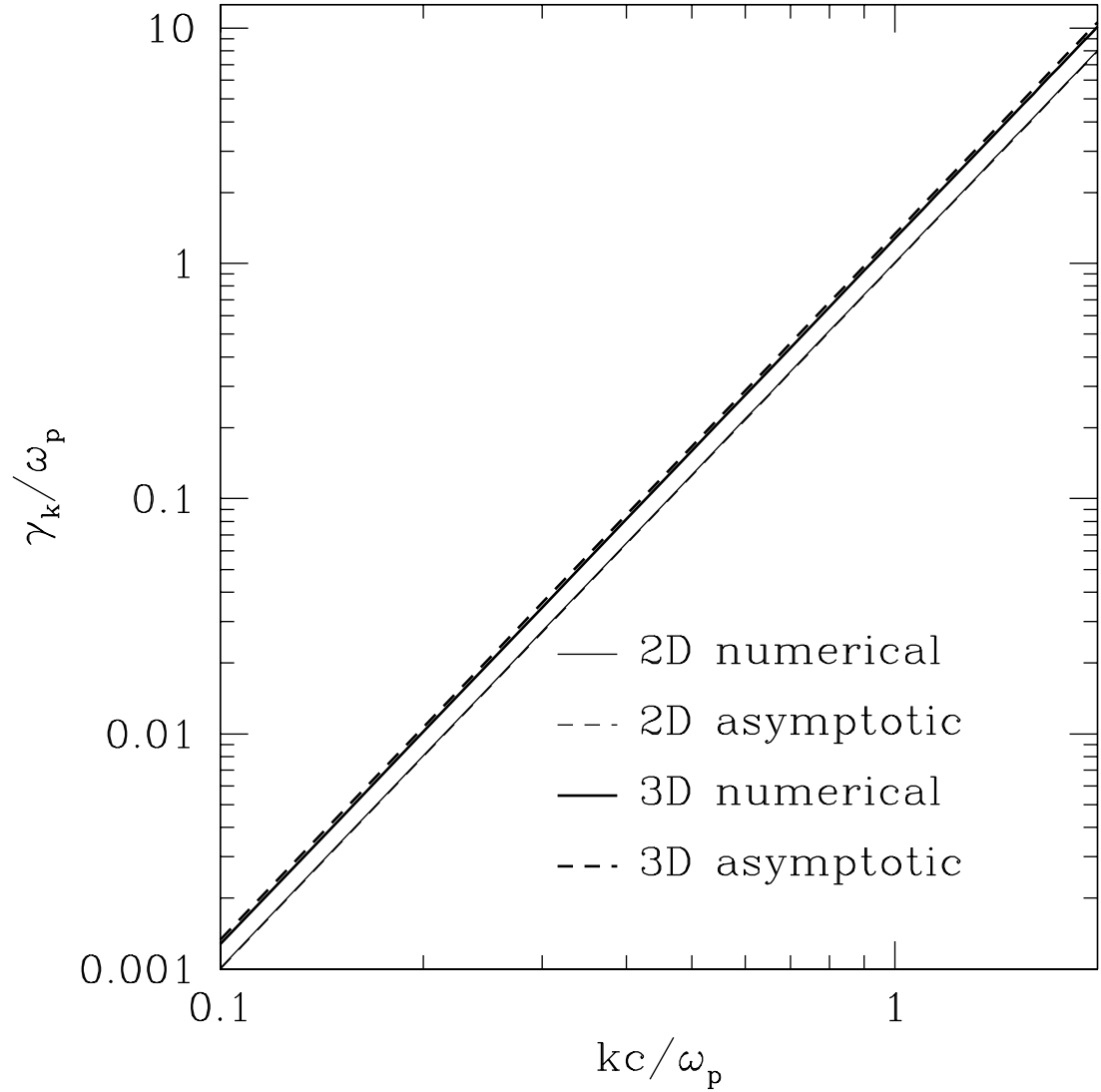}
\caption{Damping rates of magnetic field as a function of $kc/\omega_{p}$.  We plot
  the damping rates numerically computed from equation
  (\ref{eq:worked out suscept}) and (\ref{eq:2-d susceptibility}) for 3D
  (thick solid line) and 2D (thin solid line) respectively.  Also
  overplotted are the analytic forms of these expressions from
  equation (\ref{eq:gamma}) for 2D (thin dashed line) and 3D (thick
  dashed line).}
\label{fig:damping rates}
\end{figure}

Taking the general form of $4\pi\chi$ for 2D and 3D from equation
(\ref{eq:worked out suscept}) and (\ref{eq:2-d susceptibility}), we
numerically compute the damping rate from equation
(\ref{eq:app_general gamma}) for Weibel modes where $\omega_r = 0$, because of the
non-propagating nature of the downstream magnetic clumps. We
show the results in Figure \ref{fig:damping rates}. Also
plotted are the simple forms for $\gamma_k$ from equation
(\ref{eq:app_gamma}).  Though, the asymptotic forms are only valid for
$kc/\omega_{p} \ll 1$, the numerical and asymptotic results are in
excellent agreement throughout.  Thus for simplicity, we use equation
(\ref{eq:app_gamma}) (also eq.[\ref{eq:gamma}] in the main body) for
the damping rates.

%We find excellent agreement between the numerical evaluation of
%$\gamma_k$ for $\omega_r = 0$ and equation (\ref{eq:app_gamma}) (see
%Appendix A), so we use equation (\ref{eq:app_gamma}) in the rest of this
%paper.

%\clearpage

\end{document}